# Multi-Objective Problem Solving With Offspring on Enterprise Clouds

Christian Vecchiola, Michael Kirley, and Rajkumar Buyya
*Department of Computer Science & Software Engineering*
*The University of Melbourne, 3053, Carlton, Victoria, Australia*
{csve, mkirley, raj}@csse.unimelb.edu.au

## Abstract

*In this paper, we present a distributed implementation of a network based multi-objective evolutionary algorithm—called EMO—by using Offspring. Network based evolutionary algorithms have proven to be effective for multi-objective problem solving. They feature a network of connections between individuals that drives the evolution of the algorithm. Unfortunately, they require large populations to be effective and a distributed implementation can leverage the computation time. Most of the existing frameworks are limited to providing solutions that are basic or specific to a given algorithm. Our Offspring framework is a plug-in based software environment that allows rapid deployment and execution of evolutionary algorithms on distributed computing environments such as Enterprise Clouds. Its features and benefits are presented by describing the distributed implementation of EMO.*

## 1. Introduction

Many problems in science, engineering, and economics require solutions consisting of several incommensurable and possibly conflicting objectives, constraints, and/or problem parameters. Multi-objective evolutionary algorithms (MOEAs) are now a well-established population based metaheuristic used to find a set of Pareto-optimal solutions for such problems [1]. However, one of the major difficulties when applying MOEAs to real-world problems is the computational cost associated with the large number of function evaluations necessary to obtain a range of acceptable solutions. In the MOEA domain, there have only been a relatively small number of parallel models described as compared with the single objective domain (see Veldhuizen et al. [2] for a review). Recently, we introduced a novel complex network-based MOEA [3, 4], called EMO, to address the many inherent challenges when attempting to find a range of solutions, particularly for problems with a large number of objectives. Initially, we restricted the implementation to a sequential model.

In this paper, we introduce its distributed implementation by using Offspring, which is a framework we developed for distributing the execution of evolutionary algorithms. Enterprise Clouds [5] provide the required computational power to solve large optimization problems in a reasonable period. Offspring provides facilities for distributing the large computation load generated by MOEAs, by simply asking the user to define the strategy to use for coordinating the distributed execution. The primary aim of this system is to provide a friendly user environment for researchers in combinatorial optimization who do not want to be concerned about building interconnection software layers and learning underlying middleware APIs. Specifically, we provide a visual user interface that manages the execution of population based optimization algorithms, a set of APIs allowing researchers to write a plug-in for this environment quickly. The distributed version of our serial implementation of MOEA has been developed as a plug-in for this system by simply defining a strategy which: (i) coordinates the different serial executions distributed among the nodes; and (ii) applies smart migrations at the end of each of the iterations of the algorithm. Even though the infrastructure provided by Offspring is general enough to deploy a distributed implementation of any population based algorithms, the real advantage is obtained when a distributed implementation is composed by coordinating the runs of the serial implementation of the same algorithm. In that case, the distributed implementation with Offspring is obtained with a minimal coding effort, because it is only necessary to code the coordination strategy. These conditions apply to the population based metaheuristics making use of topology information to improve the quality of solutions.

The rest of the paper is organized as follows. In Section 2, we describe the related work in virtualization technologies and distributed metaheuristics. Section 3 provides a very brief introduction to population based metaheuristics and introduces the challenges in distrib-uting network based evolutionary algorithms. Section 4 and 5 describe the architecture of the distributed implementation of EMO by using Offspring. Some preliminary results are reported in Section 6. Conclusions and plans for future work follow in Section 7.

## 2. Background

The idea of providing support for distributed execution of nature inspired population based metaheuristics has been investigated with interest in the last two decades [6]. In particular, this topic has been thoroughly investigated for genetic algorithms and different parallel execution models have been devised [7, 8]. There exist a wide range of grid middleware technologies, such as Alchemi [9], Condor-G [10], the Globus Toolkit [11], and grid resource brokering technologies, such as Nimrod/G [12] and Grid Service Broker [13], that have simplified the development of distributed problem solving environments. In this work, we focus on developing the Offspring framework in .NET based Cloud Computing environments.

For what concerns distributed optimization, different solutions are now available for researchers. Nimrod/O [14] is a tool allowing running distributed optimization problems by using any Nimrod based system, such as Nimrod/G, as distribution infrastructure. Nimrod/O allows users to take advantage of different optimization algorithms (BGFS, Simplex, Divide and Conquer, and Simulated Annealing). It requires users to specify the structure of the optimization problem and the variable that needs to be minimized. ParadisEO-MOEO [15] is an object-oriented framework that provides a full featured object model for implementing distributed metaheuristics, by focusing on code reuse and efficiency. It supports MPI, Condor-G, and Globus as distributing middleware technologies. DREAM (Distributed Resource Evolutionary Algorithm Machine) [16] provides a software infrastructure and a technology for the automatic distribution of evolutionary algorithm processing. DREAM is based on a virtual machine that uses a P2P mobile agent system for distributing the computation. Other minor projects such as TEA, JDEAL, and JMETAL mostly focus on providing a good support for metaheuristic implementation and put less emphasis on the integration with distributed computing technologies.

Nimrod/O provides a technique for distributing a set of built-in optimization algorithms that is based on parameter sweeping. Offspring provides a more general approach and an extensible platform for creating distributed evolutionary algorithms. With Offspring, researchers can either define the structure of the distributed algorithm or the single computation performed on each of the nodes. These tasks cannot be performed with Nimrod/O that simply provides a technique for partitioning the problem space and distribute the computation. For these reasons, Offspring is more similar to DREAM since it provides a distribution engine making the development of distributed evolutionary algorithms straightforward. The approach used by DREAM to distribute the computation is based on mobile multi-agent systems, while Offspring relies on the Enterprise Clouds. Compared to ParadisEO-MOEO Offspring provides a smaller set of features, especially for what concerns the statistical analysis of the solutions. The API provided by ParadisEO-MOEO allows developers to virtually control any aspect of the implementation of a distributed metaheuristic. This great flexibility makes the development of a new metaheuristic not straightforward, but a good understanding of the APIs is required. The primary concern of Offspring is to provide simple and easy to use abstractions allowing researchers to compose a distributed metaheuristic by giving them the maximum freedom on the policies used to coordinate the distributed execution. As a result, the number of APIs to learn and use has been kept minimal. Moreover, another feature that distinguishes Offspring from the solutions presented is the use of Enterprise Clouds and Computational Grids.

## 3. Distributed Evolutionary Algorithms

Evolutionary algorithms are a class of population based metaheuristics [6] exploiting the concept of population evolution to find solutions to optimization problems. A population is a collection of individuals where each individual represents—or maintains information about—a specific solution of the optimization problem. The optimal solution is then found by using an iterative process that evolves the collection of individuals in order to improve the quality of the solution. Genetic Algorithms (GAs) [17] are the most popular evolutionary algorithms. They imitate the process by which nature creates new chromosomes by recombining and mutating existing chromosomes in order to generate the new population. Figure 1 describes the structure of these algorithms.

When tackling real world problems, such as those described in Handl et al. [18], the compute intensive step is the evaluation of each individual. A range of structured or parallel genetic algorithms has been proposed where the population is decentralized in some way (see Cantù-Paz [7] and Alba et al. [8] for an overview). The models may be loosely classified into one of the following four types: single-population master-slaves, multiple populations (island model), cellular

```
1. P[0] = a_1[0], …, a_n[0]     /* initialization */
2. while not stop condition S is met do
   2.1 generate a new population P[t] by using:
       mutation and recombination
   2.2 evaluate the set of solutions
   2.3 compute S
   t = t + 1
   endwhile
```

**Figure 1. Genetic Algorithms.**

(diffusion model), and hierarchical combinations. Master-slave models distribute only the evaluation phase, while multiple populations distributed the whole execution of the algorithm.

Recently, the easy access to Grid and Cloud computing infrastructures has made the deployment of hierarchical models quite common. These models compose the previously discussed models to better exploit the heterogeneity of distributed computing resources that can be found within Enterprise Clouds or Computing Grids. The execution of the evolutionary algorithm is generally divided into layers and at each of the layers a different model can be used. The most common implementation is based on a two level structure which uses a multi-population coarse grained distribution model at the first level and a master-slave or a cellular model at the second level. A recent implementation of this model has been proposed in Lim et al. [19] for genetic algorithms.

The complex network based model introduced by Kirley and Stevens [3, 4] is fundamentally a diffusion-based evolutionary algorithm. Individuals in the evolving population are mapped to the nodes of a given complex network – regular 2D lattice, small-world network, scale-free network or random network [20]. Here, the individuals interact in their local neighborhood, which is defined by the topology of the given network, and an external archive is used to store the evolved Pareto optimal front. An important feature of the algorithm was the variation in connectivity (node degree) and corresponding selection pressure across a given network. Reported results using the complex network-based model suggest that there were significant differences between the network architectures considered using the well-known ZDT benchmark multi-objective problems [21]. Significantly, relatively large population sizes are required if the inherent clustering properties of alternative complex network architectures are to be used. In order to handle these huge computation needs, a scalable hierarchical version of complex network-based model can be instantiated by employing multiple isolated populations – or islands. This model is described in Figure 2. Here, the individuals in each of the islands are mapped on to a particular topology and the evolution of the algorithm takes place separately. It is possible to tune each of the islands with the same or with different parameters settings, according to the specific distribution strategy implemented. Once the evolutionary algorithm is completed on the single computation node, the front is sent to a central coordinator node which: (i) aggregates all the fronts; (ii) performs statistical analysis; and (iii) applies migrations of individuals belonging to different populations.

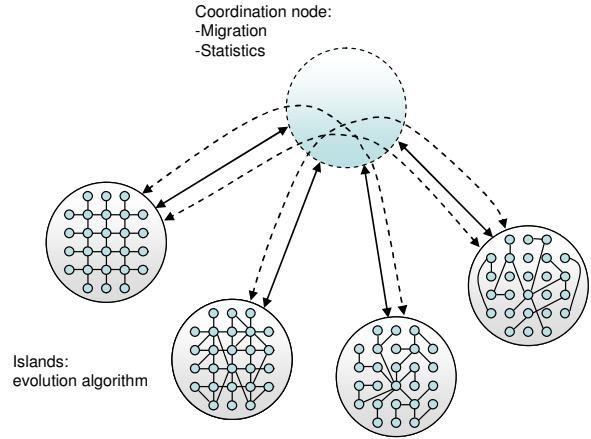

**Figure 2. The Hierarchical Complex Network Based Model.**

Such a model allows us to perform smart migrations and Offspring makes their implementation quite easy. In the next section, we will show how, given this distribution model, it is possible to rapidly prototype a distributed implementation that takes advantage of the services of Enterprise Clouds.

## 4. Architecture

### 4.1. Design Considerations

Offspring has been designed to support researchers in combinatorial optimization in quickly deploying their algorithms on a distributed computing infrastructure. In order to be effective the framework should require a minimum knowledge of Computing Grids and Enterprise Clouds from users. The requirements for this class of applications are: (i) simplicity of use; (ii) rapid development of new heuristics; (iii) support for distributed execution; and (iv) support for result analysis.

In the following, we will show how Offspring addresses these issues by describing its architecture and the main features of the system.

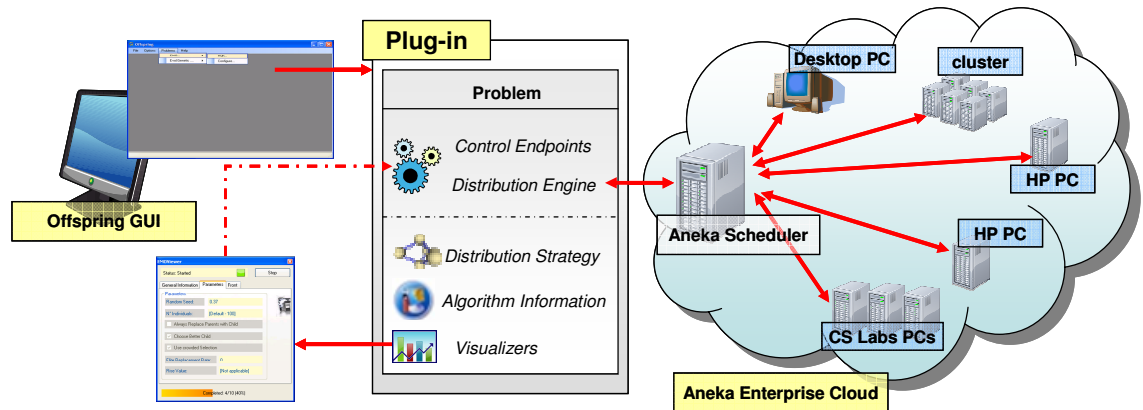

**Figure 3. System Architecture.**

### 4.2. System View

Offspring delivers to users (i) an environment through which run, monitor, and control distributed applications; (ii) a thin distribution middleware that takes care of interacting with the Enterprise Cloud; and (iii) a reference model for implementing such applications. The environment is fully customizable by using plug-ins that: (i) expose control endpoints in order to let the environment and the user visually control the execution of the algorithm; (ii) embed a distribution engine in charge of controlling the execution of the application; (iii) provide the user interface support for configuring and monitoring the execution of the application. The environment is able to load and manage multiple plug-ins and multiple applications at the same time.

Offspring provides two different integration models for building distributed applications:
- It is possible to develop a complete plug-in and then taking a finer control on how the environment interacts with the Cloud.
- It is possible to simply define distribution logic of the application, which provides to the environment the task that need to be executed at each of the iterations.

The first approach is more powerful but requires the users to know the APIs exposed by the Enterprise Cloud. The second approach makes the use of the Cloud completely transparent to the users and hence has been chosen for developing the distributed meta-heuristic discussed in this work.

### 4.3. Aneka

Offspring relies on Aneka [22] to distribute the computation of applications. Initially Aneka developed as a third generation grid technology in .NET environments. The recent advancement of Aneka introduced several new Cloud computing capabilities, such as SLA oriented resource allocation and the MapReduce programming model [23]. The main features of the platform are (i) a configurable service container hosting pluggable services for discovering, scheduling, and balancing workload; and (ii) a flexible and extensible framework/API supporting a variety of programming models such as threading, batch processing, and MapReduce. These features allow the system administrator to fine tune the installation of Aneka by carefully selecting the resources to use on each computational node. From the developer's point of view, Aneka provides a rich programming interface that allows enabling applications with support for Cloud computing quickly. Developers can choose between different execution models and select the abstraction that better fit their needs.

The distribution model of Offspring has been implemented on top of the Task Model. The Task Model is the easiest and the most general programming model supported by Aneka. It provides ready to use task classes for executing legacy code and a minimal interface for programming tasks. By using the Task Model, is possible to quickly parallelize legacy applications, or to write simple distributed applications with almost no knowledge of the distribution middleware. The use of the Aneka Task Model together with the plug-in architecture offered by Offspring allows the development of distributed metaheuristics without requiring users to know Aneka APIs: the Offspring environment takes care of interacting with Aneka, and by using the distribution logic defined by the user executes the distributed meta-heuristic.

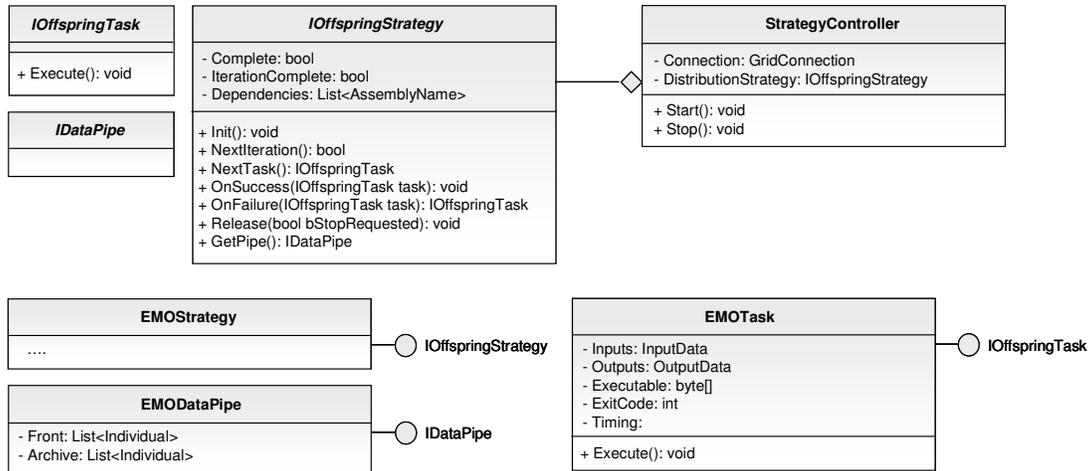

**Figure 4. Offspring Strategy and EMO++ Class Diagram.**

## 5. Implementation

### 5.1. From EMO to EMO++

The reason for porting the EMO algorithm to a distributed version is twofold. On one side, we wanted to be able to run the algorithm with a reasonable number of individuals in order to take advantage of the network-based model. On the other side we wanted to apply smart migration strategies among different population of individuals which evolved by using different network topologies. Hence, we did not need to change the structure of the algorithm but simply put a coordination strategy on top of it, which distributed the computation, evaluated results, and took decisions for the next macro iteration.

Figure 4 describes the object model exposed by Offspring for implementing population-based metaheuristics together with the implementation provided for EMO++. A strategy provides a collection of tasks that are executed on the distribution middleware by the strategy controller. In order to implement the EMO++ metaheuristic we defined a concrete class for the strategy (*EMOStrategy*), for the single task (*EMOTask*), and provided support for data visualization (*EMO-DataPipe*). In this section, we will describe how these components interact together.

### 5.2. Remote Node Execution

For what concerns the execution of EMO on the single node, there are no special requirements. It is only necessary to start the EMO algorithm with the proper input and configuration parameters and collect the results of execution. Given this, the implementation of the *EMO-Task* class consists of a very thin software layer that performs the following operations: (i) retrieves the input files and executable for the execution; (ii) starts a process and run the EMO application; (iii) waits for the termination and collects the results generated.

The amount of code required to perform these operations in C# does not exceed the body the *Execute* method of the *IOffspringTask* interface. The actual implementation provides some utility methods to monitor and control more accurately the execution but does not change the essence of the execution.

### 5.3. Implementing the Distribution Strategy

The concrete implementation of the strategy is defined in the *EMOStrategy* class that defines the distribution and coordinating logic of our metaheuristic. It provides the tasks that will be executed by means of *StrategyController* on Aneka. It controls the evolution of each of the iterations, merges the results obtained by the execution of tasks, and performs statistical analysis of data.

In Figure 5 we can see the interaction between the *StrategyController* and the *EMOStrategy* at runtime. The main execution flow is characterized by a sequence of iterations, and for each of the iterations the controller queries the strategy for a task to be executed. This execution model perfectly fits population-based metaheuristics, which are characterized by an iterative behavior. For what concerns network-based model of EMO, it is possible to distribute the computation of each of the iterations by taking advantage of the topology information connecting individuals. In this way, we can easily create a task for each group of individuals connected together. Since the execution of EMO on the single node is driven by topology information, it is not

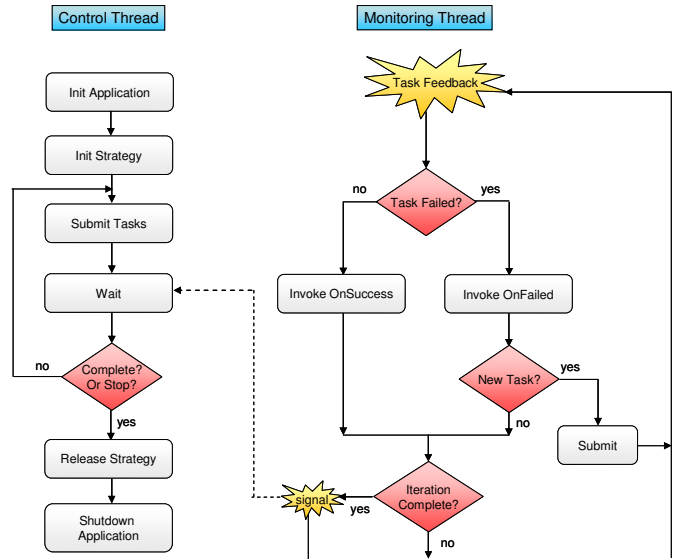
**Figure 5. Strategy Execution.**

necessary to have the complete population running on the single node but only the connected individuals.

The *StrategyController* class authenticates with Aneka by using the credential obtained from the Offspring environment, initializes the distributed application, and then the strategy. During initialization, the strategy configures the initial population, and prepares the common data for each task, such as the executable running defining the EMO algorithm. The main loop of the controller executes the iterations until the strategy does not set its *Complete* property to true. For each of the iterations the controller repeatedly asks new tasks to execute until the strategy provides a null task. The controller then puts itself in waiting mode. At the same time, a monitoring thread is responsible of collecting the tasks that completed their execution and—according to their status—of forwarding them to the strategy. The strategy merges the front with the current active front and updates statistics. For each task collected, the controller queries the strategy in order to know whether the current iteration has completed or not. If the iteration is completed the control thread is woken up and the execution proceeds to the next iteration.

This architecture concentrates within the strategy controller concurrency and distributed middleware management by keeping the definition of a strategy simple and only concerned with the implemented algorithm. The plug-in architecture previously described makes the integration with Offspring straightforward. This is done just by deploying the library containing the strategy definition and the executable of the EMO algorithm in the plug-in directory of the environment.

## 6. Performance Considerations

In order to evaluate our model, we conducted an experiment in the Computer Science department at the University of Melbourne. The Enterprise Cloud of the department is composed of one scheduler node (Dell OPTIPlex GX 270 Pentium IV 3.00 GHz, 3.25 Gb of RAM running Windows XP SP2 and .NET 2.0) and 33 computing nodes (Dell OPTIPlex GX 2f0 Pentium IV 3.40 GHz, 1.5 G of RAM running Windows XP SP2 and .NET 2.0). The Cloud resources are located into three different student laboratories connected by a 100 MB switched LAN. For what concerns the configuration of Aneka, the scheduler and the computing nodes have been set up to support the Task and the Thread programming model.

In order to test the gain obtained by the aggregate computing power of the system we executed a test suite of benchmark problems (ZDT1–ZDT6 and DLTZ1–DLTZ6) with different number of individuals and a fixed number of iterations. The numbers of individuals tested are 100, 300, 500, 1000, while the number of iterations used has been kept constant to 100. In order to compare the timing data of the serial execution with the timing data of the distributed execution we adopted the following convention: for a given serial run characterized by X individuals, we set up a distributed run that creates 10 islands with X divided by 10 individuals each.

Figure 6 shows the speed up obtained by using the distributed version compared to the serial execution. As we can notice, there is no advantage in executing the distributed version when the total number of indi-viduals is 100. The only two benchmark problems that are actually solved faster are DTLZ4 and DTLZ6, which are the most compute intensive. With a number of 300 individuals, we have a positive speed up for all the benchmark problems, but the only two problems that have super linear speed up are DTLZ4 and DTLZ6. With a number of individuals equal 500, all the DTLZ problems have a super linear speed up.

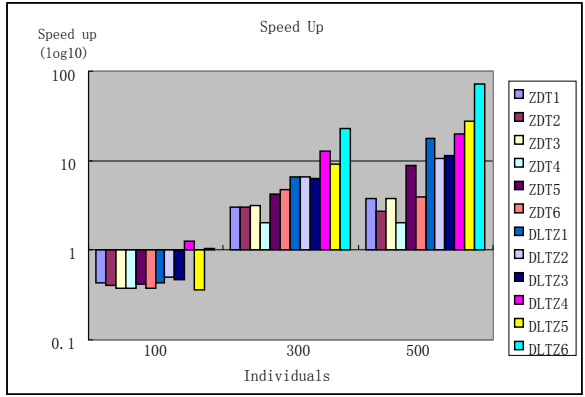

**Figure 6. Speed up Gain.**

We also measured the overhead introduced by Aneka and Offspring for each of the iterations of the algorithm. The overhead has been computed by considering the time spent from the creation of the task and the collection of its results. This value has been then compared with pure execution time of the EMO algorithm recorded on the computing node.

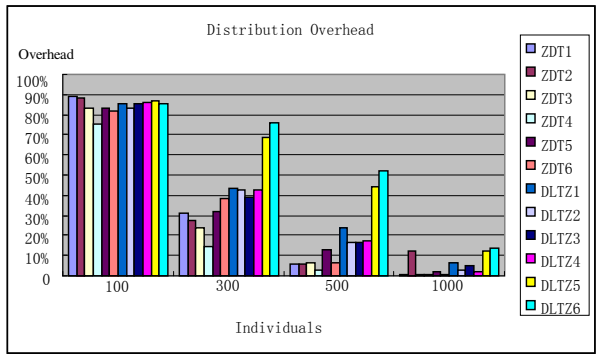

**Figure 7. Distribution Overhead.**

Figure 7 confirms the results obtained for the speed up. As it can be noticed, when the total number of individuals is 100, the average distribution overhead is about 85% of the whole execution time, and this explains why for that number of individuals we have a negative speed up. With the increase of the number of individuals, the average overhead decreases and falls below 6% of the whole computation time for populations composed by 1000 individuals.

The tests performed on the EMO++ plug-in demonstrate that there is a continuous increase in the execution speed up as the number of individuals increases. We also identified the lower bound of 100 individuals, under which the distribution infrastructure provided by Offspring does not provide any advantage. The reason why we have such a slow performance with 100 individuals is that the distribution overhead is the same order of magnitude of the execution on the local machine. In the case of network-based evolutionary algorithms this is not an issue, because this class of algorithms, in order to take advantage of the connection between individuals, requires a minimum population size that is at least ten times bigger 100 individuals.

## 7. Conclusion and Future Works

In this paper, we presented the approach proposed by Offspring for distributed multi objective evolutionary algorithms on Enterprise Clouds. The aim of Offspring is to minimize the code required to provide a distributed implementation of a population based metaheuristics without requiring the researchers to know distribution middleware APIs. Few still active frameworks have such level of abstraction and none of them relies on Enterprise Clouds. This is what motivated the authors to implement a new framework. A specific emphasis has been put in providing high degree of flexibility, ease of use, and rapid prototyping features. The most appropriate approach for delivering such support is by using plug-in architecture APIs allowing third parties to implement new solution without knowing the detail of the entire code-base.

The effectiveness of Offspring has been tested by deploying EMO++, a distributed implementation of EMO. EMO++ is a port of a real world network based—and computationally intensive—population metaheuristic. It keeps information about the connections among individuals and exploits them to evolve the population towards a better solution. This class of algorithms requires a large number of individuals in order to be effective and for this reason, it constitutes the perfect candidate for a distributed implementation and deployment with Offspring. We developed EMO++ as a strategy for Offspring and made some preliminary tests. Results show that the model proposed by Offspring is effective when there is a real need for a distributed implementation. In order to be effective network based evolutionary algorithms require at least 1000 individuals and the model proposed by Offspring provides an increasing speed up when the number of individuals is only 300. A preliminary analysis of the overhead introduced by Offspring and the Cloud middleware used shows encouraging results for large population sizes. We can then conclude that in this case the distribution infrastructure provided by Offspring does not affect the performance.

To the best of our knowledge Offspring is unique in its nature, even though is still missing some important features if compared with other solutions. In particular, support for built-in statistical analysis that is something that still need to be implemented in plug-ins.